# Bounds for effective dielectric permittivity in differential medium approximation

Vitaliy N.Pustovit
Institute of Surface Chemistry, National Academy of Sciences of Ukraine, 17 General Naumova Street, Kiev 03164, Ukraine

**Abstract**

Theoretical approach is proposed to description of dielectric properties of matrix disperse systems which consists of dielectric matrix with embedded in metallic inclusions. On the basis of effective differential medium approximation the analytical expressions are obtained for the effective dielectric permittivity $\widetilde{\varepsilon}$ of the matrix disperse system with inclusions of spherical and ellipsoidal shape. The analysis of limits of possible values of the real and imaginary parts of $\widetilde{\varepsilon}$ is carried out depending on system parameters.

[*] e-mail: pustovit@ccmsi.us

## 1. Introduction

Under theoretical study of processes of interaction of electromagnetic radiation with matrix disperse systems (MDS), which represent a continuous matrix (as a rule dielectric one) with the imbedded inclusions of the various forms and nature, the effective medium approximations widely used. The essence of this method consists in that the MDS with distributed values of dielectric permittivity of a matrix $\varepsilon_1$ and inclusions $\varepsilon_2$ is replaced by continuous medium with effective dielectric permittivity $\widetilde{\varepsilon}$. It is depend both on $\varepsilon_1$ and $\varepsilon_2$, and on value $f = V/V_0$ (where $V_0$ is a volume engaged by inclusions, $V$ is a total volume of system) and its statistical distribution in a matrix. Such approximation well correlated to the experiment in a case, when wavelength of the incident electromagnetic radiation, interacting with MDS, is significant larger than characteristic sizes of inclusions and mean distance between them (long-wave approximation). There are many methods of calculation of $\widetilde{\varepsilon}$ for similar systems. The review of the literature concerning this problem can be found in [1-2]. In the present paper the calculation of effective dielectric permittivity $\widetilde{\varepsilon}$ is carried out under the large concentration of metallic inclusions of the spherical and ellipsoidal shape in approximation of differential effective medium (DEM) [3-5]. The estimation of possible limits of values $\widetilde{\varepsilon}'$ and $\widetilde{\varepsilon}''$ ($\widetilde{\varepsilon} = \widetilde{\varepsilon}' + i\widetilde{\varepsilon}''$) is obtained with use of common methods of electrodynamics in inhomogeneous medium.

## 2. The MDS effective dielectric permittivity in DEM approximation

A correct calculation of effective dielectric permittivity ($\widetilde{\varepsilon}$) of MDS under the large volume fraction of inclusions represents a very complicated problem. One of methods of $\widetilde{\varepsilon}$ calculation is a method of differential effective medium (DEM) [3-4].

Generally DEM follows from the average field approximation (Bruggeman's approximation [6]). In case of inclusions of ellipsoidal shape the Bruggeman's approximation for calculation of $\widetilde{\varepsilon}$ has a form:

$$F(f, \varepsilon_1, \varepsilon_2, \widetilde{\varepsilon}) = \frac{(1-f)(\widetilde{\varepsilon} - \varepsilon_1)}{(1-L)\widetilde{\varepsilon} + L\varepsilon_1} + \frac{f(\widetilde{\varepsilon} - \varepsilon_2)}{(1-L)\widetilde{\varepsilon} + L\varepsilon_2} + \frac{4(1-f)(\widetilde{\varepsilon} - \varepsilon_1)}{(1+L)\widetilde{\varepsilon} + (1-L)\varepsilon_1} + \frac{4f(\widetilde{\varepsilon} - \varepsilon_2)}{(1+L)\widetilde{\varepsilon} + (1-L)\varepsilon_2} = 0 \quad (1)$$

where $L$ is a depolarization factor along the large semi-axis ($L_1 = L$; $L_2 = L_3 = \frac{1-L}{2}$), $\varepsilon_2$ and $\varepsilon_1$ are dielectric permittivity of inclusions and matrix, respectively. Consider first the variation of the effective dielectric permittivity at the expense of adding of a small portion of particles of the second phase with relative volume $\Delta f$. This change will be:

$$\Delta \tilde{\varepsilon} = -\frac{\left(\partial F/\partial f\right)\big|_{f=0}}{\left(\partial F/\partial \tilde{\varepsilon}\right)\big|_{f=0}} \Delta f = g(\varepsilon_1, \tilde{\varepsilon})\Delta f \qquad (2)$$

After differentiation Eq.(1) we find:

$$g(\tilde{\varepsilon}, \varepsilon_1) = \frac{\tilde{\varepsilon}}{3}(\varepsilon_1 - \tilde{\varepsilon})\left[\frac{1}{(1-L)\tilde{\varepsilon} + L\varepsilon_1} + \frac{4}{(1+L)\tilde{\varepsilon} + (1-L)\varepsilon_1}\right] \qquad (3)$$

In the result of addition to the effective medium $\tilde{\varepsilon}$ of $\Delta f$ portion of particles of the second phase, $f\Delta f$ particles of the second phase will be replaced. Therefore, true change of a fraction of particles of the second phase in a new composite will be: $\Delta f_0 = (1-f)\Delta f$, i.e. in the relation (2) it is necessary $\Delta f \to \frac{\Delta f}{1-f}$. In view of it, the relation (2) will have a form:

$$\frac{d\tilde{\varepsilon}}{g(\tilde{\varepsilon}, \varepsilon)} = \frac{df}{1-f} \qquad (4)$$

The solution of the Eqs. (3) - (4) with the initial condition $\tilde{\varepsilon} = \varepsilon_1$ at $f = 0$ gives:

$$1 - f = \left(\frac{\varepsilon_1}{\tilde{\varepsilon}}\right)^a \left(\frac{\varepsilon_1 + \gamma\varepsilon_2}{\tilde{\varepsilon} + \gamma\varepsilon_2}\right)^b \left(\frac{\tilde{\varepsilon} - \varepsilon_2}{\varepsilon_1 - \varepsilon_2}\right) \qquad (5)$$

where $a = \frac{3L(1-L)}{1+3L}$; $b = \frac{2(1-3L)^2}{(1+3L)(5-3L)}$; $\gamma = \frac{1+3L}{5-3L}$

Thus, for the case of particles of spherical form at $L = \frac{1}{3}$ and $a = \frac{1}{3}$, $\gamma = \frac{1}{2}$, $b = 0$, this formula takes a form [4]:

$$1 - f = \left(\frac{\varepsilon_1}{\tilde{\varepsilon}}\right)^{1/3}\left(\frac{\tilde{\varepsilon} - \varepsilon_2}{\varepsilon_1 - \varepsilon_2}\right) \qquad (6)$$

Let us to analyze the case of symmetric replacement $f \Leftrightarrow \psi = (1-f)$ in the formula (1). Really, such replacement can take place when considering the dielectric in MDS as inclusions, and metal particles as a matrix. The physical sense of the formula does not alter from such replacement. Then from (1) we obtain:

$$f = \left(\frac{\varepsilon_2}{\widetilde{\varepsilon}}\right)^a \left(\frac{\varepsilon_2 + \gamma\varepsilon_1}{\widetilde{\varepsilon} + \gamma\varepsilon_1}\right)^b \left(\frac{\widetilde{\varepsilon} - \varepsilon_1}{\varepsilon_2 - \varepsilon_1}\right) \qquad (7)$$

For the case of particles of spherical shape at $L = \frac{1}{3}$ the given formula has a form [3]:

$$f = \left(\frac{\varepsilon_2}{\widetilde{\varepsilon}}\right)^{1/3} \left(\frac{\widetilde{\varepsilon} - \varepsilon_1}{\varepsilon_2 - \varepsilon_1}\right) \qquad (8)$$

From the Eqs. (6) and (8) it follows the relation for calculation of $\widetilde{\varepsilon}$:

$$\frac{1-f}{\varepsilon_1^{1/3}} + \frac{f}{\varepsilon_2^{1/3}} = \frac{1}{\widetilde{\varepsilon}^{1/3}} \qquad (9)$$

The Eq. (9) belongs to a common class of relations of a kind

$$\widetilde{\varepsilon}^k = f\varepsilon_2^k + (1-f)\varepsilon_1^k \qquad (10)$$

which, as shown in [7], at $|k| \leq 1$ completely satisfies to the theorems on the limits of possible values of the effective dielectric permittivity for MDS and statistical mixtures [7-12].

Let us to consider now a behavior of $\widetilde{\varepsilon}$ depending on parameters of the system. For the analysis of behavior of effective dielectric permittivity in approximation DEM we shall use Eq.(7), where

$$\widetilde{\varepsilon} = \varepsilon + i\chi; \quad \varepsilon_m = \varepsilon_0 + i\chi_0; \quad \varepsilon = \varepsilon_1 + i\chi_1 \qquad (11)$$

and, accordingly:

$$\begin{aligned}
\widetilde{\varepsilon} &= (\varepsilon^2 + \chi^2)^{1/2} \exp\left(i \cdot arctg \frac{\chi}{\varepsilon}\right) \\
\varepsilon &= (\varepsilon_1^2 + \chi_1^2)^{1/2} \exp\left(i \cdot arctg \frac{\chi_1}{\varepsilon_1}\right) \\
\varepsilon_m &= (\varepsilon_0^2 + \chi_0^2)^{1/2} \exp\left(i \cdot arctg \frac{\chi_0}{\varepsilon_0}\right)
\end{aligned} \qquad (12)$$

Thus, we have an opportunity to divide in relation (7) the real and imaginary parts.

a) For a real part we have:

$$f^2 = \left[\frac{\varepsilon_1^2 + \chi_1^2}{\varepsilon^2 + \chi^2}\right]^a \left[\frac{(\varepsilon_1 + \gamma\varepsilon_0)^2 + (\chi_1 + \gamma\chi_0)^2}{(\varepsilon + \gamma\varepsilon_0)^2 + (\chi + \gamma\chi_0)^2}\right]^b \left[\frac{(\varepsilon - \varepsilon_0)^2 + (\chi - \chi_0)^2}{(\varepsilon_1 - \varepsilon_0)^2 + (\chi_1 - \chi_0)^2}\right] \qquad (13)$$

b) For an imaginary part we have:

$$a \cdot arctg\left[\frac{\varepsilon_1 \chi - \chi_1 \varepsilon}{\varepsilon_1 \varepsilon + \chi_1 \chi}\right] = b \cdot arctg\left[\frac{(\chi_1 + \gamma\chi_0)(\varepsilon + \gamma\varepsilon_0) - (\chi + \gamma\chi_0)(\varepsilon_1 + \gamma\varepsilon_0)}{(\varepsilon_1 + \gamma\varepsilon_0)(\varepsilon + \gamma\varepsilon_0) + (\chi_1 + \gamma\chi_0)(\chi + \gamma\chi_0)}\right] +$$
$$+ arctg\left[\frac{(\chi - \chi_0)(\varepsilon_1 - \varepsilon_0) - (\chi_1 - \chi_0)(\varepsilon - \varepsilon_0)}{(\varepsilon - \varepsilon_0)(\varepsilon_1 - \varepsilon_0) + (\chi - \chi_0)(\chi_1 - \chi_0)}\right] \quad (14)$$

From these general expressions it is very complicated to define behavior of effective dielectric permittivity of the system. Therefore, it is necessary to make some of approximation. As an example, consider low-frequency limit at $\omega \to 0$, when $\chi \gg \varepsilon$, and assuming that $\chi_0 = 0$ for the matrix. Then, from the Eqs. (13) and (14) we obtain, respectively:

a)
$$\operatorname{Im}\tilde{\varepsilon}(\omega \to 0) = \chi = \chi_1 f^{-1/\beta}, \quad (15)$$

where $\beta = a + b - 1$

b) In approximation $arctg(x) \approx x$

$$\beta\left(\frac{\varepsilon_1}{\chi_1} - \frac{\varepsilon}{\chi}\right) = \alpha\varepsilon_0\left(\frac{1}{\chi} - \frac{1}{\chi_1}\right), \quad (16)$$

where $\alpha = b\gamma + 1$.

After substitution of (15) into expression (16), we have:

$$\operatorname{Re}\tilde{\varepsilon}(\omega \to 0) = \varepsilon = f^{-1/\beta}\left[\varepsilon_1 + \frac{\alpha}{\beta}\varepsilon_0\right] - \frac{\alpha\varepsilon_0}{\beta}. \quad (17)$$

The condition of $\operatorname{Re}\tilde{\varepsilon}$ sign changing, corresponded to the metal-dielectric optical transition phenomena [3,5] we find from the equation $\operatorname{Re}\tilde{\varepsilon} = 0$, which produces:

$$f^* = \left(1 + \frac{\varepsilon_1 \beta}{\alpha\varepsilon_0}\right)^\beta. \quad (18)$$

3. **Estimation of limit values $\tilde{\varepsilon}$ for disperse systems**

The following inequalities (Wiener relations [9]) for $\tilde{\varepsilon}$ are valid for real $\varepsilon_2$ and $\varepsilon_1$ for the two-component system:

$$\varepsilon_- < \tilde{\varepsilon} < \varepsilon_+ \quad (19)$$

and

$$\frac{1}{\varepsilon_-} = \frac{f_1}{\varepsilon_1} + \frac{f_2}{\varepsilon_2} \quad (20)$$

$$\varepsilon_+ = f_1\varepsilon_1 + f_2\varepsilon_2 \quad (21)$$

where $f_1$ and $f_2$ are volume fractions $(f_1 + f_2 = 1)$; $\varepsilon_1$ and $\varepsilon_2$ are dielectric permittivities of medium components; $\varepsilon_+$ is a dielectric permittivity of layered dielectric in electrical field parallel to the border of the unit of layers; $\varepsilon_-$ is a dielectric permittivity of layered dielectric in electrical field perpendicular medium to the plane. Later in the paper [10] the estimations of a type of Eq.(19) were improved and generalized on the case of complex $\varepsilon_1$ and $\varepsilon_2$ [11-12]. In case of real $\varepsilon_1$ and complex $\varepsilon_2$ the limits of area of possible values of $\widetilde{\varepsilon}'$ и $\widetilde{\varepsilon}''$ ($\widetilde{\varepsilon} = \widetilde{\varepsilon}' + i\widetilde{\varepsilon}''$) are given by set of relations (20) and (21) [8], which after introduction of variables $x = \dfrac{\widetilde{\varepsilon}}{\varepsilon_1}$ and $t = \dfrac{\varepsilon_2 - \varepsilon_1}{\varepsilon_1}$ will have a form:

$$x'' = \frac{t''}{t'}(x' - 1) \qquad (22)$$

$$(x' - a)^2 + (x'' - b)^2 = R^2 \qquad (23)$$

$$a = \frac{1}{2}; \quad b = \frac{1}{2t''}(t' + |t|^2); \quad R = \sqrt{a^2 + b^2}$$

where $x'$ and $x''$ ($x = x' + ix''$) are real and image parts of $x$; $t'$ and $t''$ ($t = t' + it''$) are the same values for $t$. When obtaining Eq.(22) and (23) from (20) and (21) the value $f$ was excluded. The straight line Eq.(22) in a plane variables of $(x', x'')$ passes through a point $A$ ($x' = 1; x'' = 0$ corresponds to $\varepsilon_1$) and $B$ ($x' = 1 + t'; x'' = t''$ corresponds to $\varepsilon_2$). Through the same points also passes the circle Eq.(23), which passes, besides, through the coordinate origin. As a result of crossing Eq.(22) and Eq.(23) the segment of the permitted values $x'$ and $x''$ (Fig.1) is formed. It should be noted that $x'$ and $x''$, as it follows from (20) and (21) are parametrically depend on $f$ ($f = f_2; f_1 = 1 - f$).

In the paper [7] the improved estimations of limits of permitted values of $\widetilde{\varepsilon}$ are obtained. They look like Eq.(10) and Eq.(11) from [7]:

$$(\varepsilon_1 - \widetilde{\varepsilon})(\varepsilon_1 * - \widetilde{\varepsilon}*) + f\frac{(\varepsilon_1 - \varepsilon_2)(\varepsilon_1 * - \varepsilon_2 *)(\varepsilon_1 * \widetilde{\varepsilon} - \varepsilon_1 \widetilde{\varepsilon}*)}{\varepsilon_1 \varepsilon_2 * - \varepsilon_2 \varepsilon_1 *} \leq 0 \qquad (24)$$

$$(\varepsilon_2 - \widetilde{\varepsilon})(\varepsilon_2 * - \widetilde{\varepsilon}*) + (1 - f)\frac{(\varepsilon_2 - \varepsilon_1)(\varepsilon_2 * - \varepsilon_1 *)(\varepsilon_2 * \widetilde{\varepsilon} - \varepsilon_2 \widetilde{\varepsilon}*)}{\varepsilon_2 \varepsilon_1 * - \varepsilon_1 \varepsilon_2 *} \leq 0 \qquad (25)$$

These relations easily can be written down in variable $t$:

$$(x' - a_1)^2 + (x'' - b_1)^2 = R_1^2$$

$$a_1 = 1; \quad b_1 = \frac{f|t|^2}{2t''}; \quad R_1 = b_1 \qquad (26)$$

$$(x'-a_2)^2 + (x''-b_2)^2 = R_2^2$$

$$a_2 = \left(1+t'+\frac{1-f}{2}|t|^2\right); \quad b_2 = \frac{(1-f)(1+t')|t|^2}{2t''}; \quad R_2 = \frac{(1-f)|1+t'||t|^2}{2t''} \qquad (27)$$

As a result of intersection circles (26) and (27) the small segment is formed, which tops lay on a straight line (22) and circle (23), whereas its position in sector depends on $f$. One can show [7], that relations for calculation of $\tilde{\varepsilon}$ in a form (10) under $|k| \leq 1$ satisfy to the given above restrictions. It should be noted that a number of known relations belong to the condition (10). When $k = \pm 1$ we have (20) and (21), when $k = 1/3$ relation of L.D.Landau [13], and when $k = -1/3$ relation (9), etc. The case $k = 0$ is a special and $\tilde{\varepsilon}$ can be found from relation of Lichtenecker [14]:

$$\ln \tilde{\varepsilon} = f \ln \varepsilon_2 + (1-f) \ln \varepsilon_1 \qquad (28)$$

The relation (28) also satisfies to above-mentioned restrictions. On Fig. 1 the area of allowed values of $\tilde{\varepsilon}'$ and $\tilde{\varepsilon}''$ is shown for the case $f = 0.5$ at $\varepsilon_1 = 2$; $\varepsilon_2 = 5+10i$. The points in the figure correspond to values of $\tilde{\varepsilon}$ calculated by the formulas:

point "1" is Maxwell Garnett approximation [2]

$$\frac{\tilde{\varepsilon} - \varepsilon_1}{\tilde{\varepsilon} + 2\varepsilon_1} = f \frac{\varepsilon_2 - \varepsilon_1}{\varepsilon_2 + 2\varepsilon_1} \qquad (29)$$

point "2" is the average field approximation [6]

$$(1-f)\frac{\tilde{\varepsilon} - \varepsilon_1}{\varepsilon_1 + 2\tilde{\varepsilon}} + f \frac{\tilde{\varepsilon} - \varepsilon_2}{\varepsilon_2 + 2\tilde{\varepsilon}} = 0 \qquad (30)$$

point "3" is the formula (8), point "4" is the formula (9), point "5" is the formula of L.D.Landau [13]:

$$\tilde{\varepsilon}^{1/3} = f \varepsilon_2^{1/3} + (1-f)\varepsilon_1^{1/3} \qquad (31)$$

All calculated values of $\tilde{\varepsilon}$ lay in segment area, i.e. in the field of permitted values. On Fig.2 the dependence $\tilde{\varepsilon}$ on $f$ is shown in the case of real values $\varepsilon_1 = 2$; $\varepsilon_2 = 40$ calculated by relations (29) - (31) and (8) - (9) for the value $L = \frac{1}{3}$. From figure it is visible, that all curves lay in area limited straight line Eq.(21) and hyperbole Eq.(20):

$$\chi_+ = 1 + ft \tag{32}$$

$$\chi_- = \frac{1+t}{(1-f)(1+t)+f} \tag{33}$$

## 4. Discussion and conclusions

Proceeding approximation for differential effective media (DEM) in a low-frequency limit $\omega \to 0$, the condition of optical transition metal - dielectric for the dielectric media contained inclusions of the elliptical shape is obtained. In opposite to results received by application of Bruggeman's theory [16], it is evident dependence of the point of transition metal - dielectric (18) on the dielectric permittivity of a matrix $\varepsilon_0$. As an example of application formula (18) in the case of spherical particles, when $L = \frac{1}{3}$; $\nu = 0.1$; $\beta = -\frac{2}{3}$; $\alpha = 1$; $\varepsilon_0 = 10$, we have transition point at $f^* = 0.26$, that is much larger than predicted by Bruggeman's theory for the particles with elliptical shape [16]. The transition at the point $f = f^*$ has received the name of the optical transition, near to which at $f < f^*$ we have $\mathrm{Re}\,\tilde{\varepsilon} > 0$ - «dielectric» behavior of dielectric function, and at $f > f^*$, $\mathrm{Re}\,\tilde{\varepsilon} < 0$ - «metal» behavior of dielectric function. Near to concentrations $f = f^*$ the optical properties of such metallic composites practically do not depend on frequency $\omega$ of the electromagnetic radiation, that interacts with them.

The received relations (5) - (9) for calculation of $\tilde{\varepsilon}$ in DEM approximation for the case of spherical, and ellipsoidal inclusions do not contradict conditions of restrictions Eq.(23) and (26) - (27) for values $\tilde{\varepsilon}'$ and $\tilde{\varepsilon}''$. However there are several moments on which we would like to discuss. When restricting with a case of spherical inclusions, as follows from relations (29) and (30), the structure of inclusions (spherical shape) is present in expressions (29) and (30) through multipliers $\frac{\varepsilon_2 - \varepsilon_1}{\varepsilon_2 + 2\varepsilon_1}$, $\frac{\tilde{\varepsilon} - \varepsilon_1}{\varepsilon_1 + 2\tilde{\varepsilon}}$, $\frac{\tilde{\varepsilon} - \varepsilon_2}{\varepsilon_2 + 2\tilde{\varepsilon}}$. These multipliers are determined within accuracy of the value **a** (where **a** is the particle radius) polarizability of a particle. It should be noted that the generalization of relations (29) and (30) to case of ellipsoidal inclusions can be found in work [1]. The particles through the expression structures (8) are present in relations of DEM approximation Eq.(5) partially, and in the case of approximations Eqs.(9), (10) and (28) they are absent at all. The importance of account of inclusions structure is especially seen processes of absorption of microwave electromagnetic radiation in matrix systems with metal inclusions. Thus, in the case of inclusions of the spherical shape under small $f$ the peak of absorption on a frequency close to the surface plasmon frequency of inclusion $\sim \dfrac{\omega_p}{\sqrt{3}}$ [2] is found out.

The value $\omega_s$ can be found from a condition $\varepsilon'_2 + 2\varepsilon_1 = 0$, which does not enter in any way into relations of DEM approximation. It seems, that relations (5) - (8), very well describe electrodynamics properties of various statistical mixtures in low-frequency area under large values of $f$. Their using makes attainable description of electrodynamical and other properties of the dump porous systems (sand, ground, rocks, etc…). Proceeding from the relations (5) - (8), the theoretical explanation of the experimentally found Archi law [15] for effective conductivity $(\tilde{\sigma})$ of such rocks was obtained:

$$\tilde{\sigma} = \sigma_2 f^{3/2} \qquad (34)$$

where $\sigma_2$ is a conductivity of the conducted fraction.

Thus, despite of limitations of relations (9), (28) - (31), they can be used for calculation of electrodynamics properties two-component systems practically in all interval variation of the value $f$ $(0 \le f \le 1)$.

**References**

[1] R.Landauer, First conference on the Electrical transport and optical properties of heterogeneous media, Ohio State University- AIP Conf. Proc. **40**, (1978), 2.

[2] C.F. Boren and P.R. Hafmen, Absorption and Scattering of Light by Small Particles, *Wiley*, New York, (1983).

[3] P.N.Sen, G.Scala, and M.H.Cohen, Geophysics, **46**, 5, (1981), 781.

[4] T. Hanai, Emulsion science, P.Sherman Ed., (1968), New York, Academic Press, 256.

[5] P.Sheng, Geophysics, **56**, 8, (1991), 1236.

[6] D.A.G. Bruggeman, Ann. Physik., **24**, (1935), 636.

[7] V.A. Antonov, V.I. Pshenizin, Optics and spectroscopy, **30**, 2, (1981), 362.

[8] D.J.Bergman, D.Stroud, Solid State Physics, **46**, (1992), 148.

[9] O.Wiener, Abh. Sachs, Acad.Wiss. Leipzig Math., **32,** Naturwiss, Kl. (1912),509.

[10] Z.Hahsin and S.Shtrikman, J.Appl.Phys., **33,** (1962),3125.

[11] G.W.Milton, Appl.Phys. A, **26**, (1981), 207.

[12] D.J.Bergman, in "Homogenization and Effective Media", Springer-Verlag, New York, (1986), pp.37.

[13] L.D.Landau and E. Lifshitz, Electrodynamics of continuous medium, -M , Nauka, (1960), pp.248.

[14] K. Lichtenecker, Physik. Z., **25,** (1924), 169.

[15] G.E. Archie., Trans.AIME , **146**, (1942), 54.

[16] L.G.Grechko, V.N.Pustovit, V.V.Motrich, S.N. Shostak, Physics of condensed and high molecular matter, **6**, (1998), Rovno, 52.

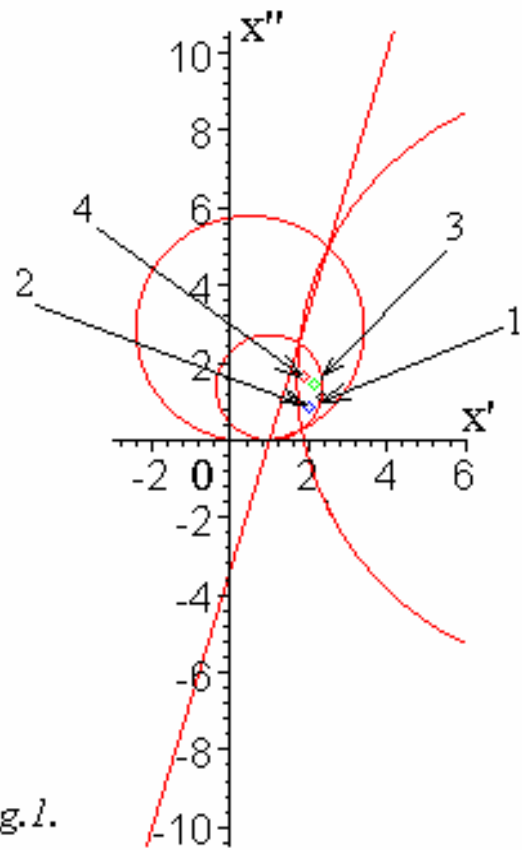

**Fig.1.** The limits of real and imaginary part of effective dielectric permittivity in the case of inclusions of spherical shape at $f = 0.5$ and $\varepsilon_1 = 2$, $\varepsilon_2 = 5 + 10i$

1). Maxwell Garnett approximation Eq.(29).
2). The approximation Eq.(9).
3). Landau approximation Eq.(31).
4). DEM approximation Eq.(6).

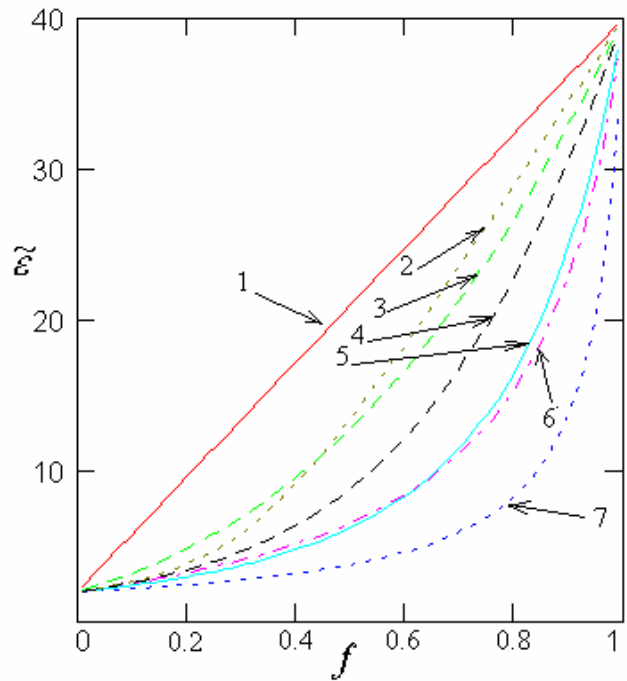

**Fig.2.** Dependence of dielectric permittivity $\widetilde{\varepsilon}$ of MDS on the volume fraction of inclusions $f$, calculated by the different effective medium approximations at $\varepsilon_1 = 2$, $\varepsilon_2 = 40$

1) The upper limit of value $\widetilde{\varepsilon}$ Eq. (21)
2) Bruggeman's approximation Eq. (30)
3) Landau approximation Eq. (31)
4) DEM approximation Eq.(6)
5) The approximation Eq.(9)
6) Maxwell Garnett approximation Eq.(29)
7) The down limit of value $\widetilde{\varepsilon}$ Eq. (20)